# Routing Protocols Performance In Mobile Ad-Hoc Networks Using Millimeter Wave

Mustafa S. Aljumaily

Department of Electrical Engineering and Computer Science
The University of Tennessee, Knoxville, TN, USA

## Abstract

*Self-Organized networks (SONs) have been studied for many years, and have attracted many researchers due to their substantial applications. Although the performance of such networks in the lower band networks (sub-6 GHz band frequencies) has been well studied, there are only sparse studies on SON in higher frequency bands, such as the millimeter wave (mmWave) band ranges between 28GHz and 300GHz. mmWave frequencies have attracted many researchers in the past few years because of its unique features and are now considered as an important part of the next generation of wireless communications namely (5G).In this paper, we study the performance of some well-known routing protocols in the case of mmWave Mobile Ad hoc Networks (MANET) using the ns-3 mmwave module that was developed recently. SONs are within the goals for the next release of the 3GPP New Radio (NR) standardization process (Release-16) for the 5G, which makes the study of the behavior of such frequency bands for these networks an important activity towards achieving such goal. Mathematical and simulation results show a great improvement in the routing protocols delivery rates and power consumption when using mmWave compared to the sub-6GHz band frequencies.*

## 1. Introduction

Mobile Ad-hoc Networks (MANETs) have been studied for many years and they are the networks formed solely from mobile User Equipement (EU) that are cooperating to exchange data in an Infrastructure-less environment [4]. MANET can be used for many applications include the tactical edge operations, disastrous areas, and in the congested environments like campuses and stadiums where many users are willing to exchange information directly with each other or using others' devices are routers. Fifth generation (5G) of wireless communications is intended to provide much higher data rates and much lower end-to-end over-the-air (OTA) latency [7]. Some prospective applications for the 5G (besides the traditional cellular communications) are the wireless virtual reality (VR), Augmented Reality (AR), Device to Device (D2D) communications in the network edges, and the autonomous vehicles in the Vehicular Ad-hoc Networks (VANET) [1], [7], which can be part of an infrastructural or infrastructure-less networks. Millimeter wave frequencies (mmWave) are expected to have a major role in the 5G standards [2]. They have their advantages of huge available bandwidth (several GHz) and reduced delay, while they also have some limitations that are related to the limited transmission range, and the need for transmitting narrow beams to cover larger distances. This work is intended to test the performance of a mobile ad-hoc network that consists of only mmWave user equipment (UE) without eNodeB (or gNodeB as suggested recently by the 3GPP [3]). The delivery ratio of data that is transmitted between any two mobile nodes in the network is of a significant importance for the feasibility and stability of the mmWave MANET applications. This type of networks with dynamic topology (because of mobility and lack of Infrastructure) is unsuitable for traditional end-to-end routing algorithms, and that is why many MANET routing protocols have been proposed to control forwarding data from any node to any other node in the multi-hop MANET network [4].

Traditional routing protocols for ad-hoc networks are usually dependent on the broadcast nature of wireless signals in the sub-6 GHz band [8]. With the mmWave's directional antennas and beam forming, this argument about wireless signals is no longer true. To compensate for such a shortage, mmWave devices use many approaches to scan the entire environment around them like beam sweeping, random beam forming (RBF), et al. [20] and send narrow directional beams towards the intended destination nodes to mitigate the large propagation path loss.

Our contributions can be summarized as:

• We list some of the well known and the recently suggested channel models and the corresponding path propagation loss and the expected received power for each case.

• We investigate the effect of multiple mmWave channel models (for different environments) that have been measured in recent years [12], [14] on the performance (delivery rate, error rate, energy efficiency, ...etc.) of some well-known MANET routing protocols in the literature.

• We prove with the help of simulation that using mmWave frequency bands (ex. 28GHz) with the traditional routing protocols for mobile ad hoc networks can improve the performance and reduce the energy consumption to a large extent.

The rest of this paper will be organized as follows. Section 2 will show some related work to the MANET generally and the mmWave in MANET. Section 3 will explain some features of the used protocols and how their performance is expected to improve by using mmWave frequencies. In sections 4 and 5, we will talk about the performance evaluation metrics and the results of performance evaluation respectively. Finally, section 6 will conclude the paper and give some ideas about our future work in this field.

## 2. RELATED WORK

Mobile Ad hoc Networks (MANET) have been attracting a lot of research attention for many years now. Many routing protocols were proposed for these networks, and the ones that are analyzed in this work [9]–[11] are among the most famous ones. Besides interesting in providing efficient data forwarding protocols in these networks, ensuring link availability and network stability have been studied as well in [16], [17]. Optimizing routes to achieve the ergodic rate density (ERD) in each link has been proven in [18], though they considered as the upper bound that can be achieved and some sub-optimal and more realistic protocols have also be presented [18].

Many geographic based routing protocols have been for such mobile networks as in [24], [25]. In [24], a parallel routing protocol (PRP) for MANET was proposed, where multiple data packets over disjoint paths can be routed simultaneously. Though they assume that each node in the network can maintain updated information about its own location in the virtual grid of the network using GPS which is not always available for such nodes in reality. In [25], several different routing protocols for MANET were analyzed and their performance was compared for special MANET networks that are used for video streaming with all its special requirements. It was found that video streaming is possible for such networks using the traditional routing protocols with acceptable quality [25]. Although, they show that the performance of any routing protocol varies depending on the network scenario and the type of video traffic used. In [21], they use a stochastic geometry approach to characterize the one-way and two-way communication characteristics and especially the Signal to Interference Ratio (SIR), and Interference to Noise Ratio (INR) distributions of a mmWave ad-hoc network with directional antennas with random blockage model, and ALOHA channel access. Other work that tried to utilize the mmWave in MANET was done in [19] where they propose an Optimal Geographic Routing Protocol (OGRP) and a directional Medium Access Control (MAC) protocol for MANET with small range and using directional antennas. Though, that work did not analyze or compare the performance of the suggested protocol with other protocols that are normally working in Wi-Fi networks. Other than that, there have been no efforts to study the effect of using mmWave on the performance of the MANET in the literature.

## 3. MANET ROUTING ALGORITHM WITH MM WAVE

### 3.1 System Model

Mobile ad hoc networks are consist of many User Equipment (UE) that are capable of transmitting and receiving directly from each other without the need for network infrastructure [4]. Each UE can be a transmitter, relay, or receiver nodes in any data transmission, and each UE has a specific transmission range that depends on the transmission power, a frequency band used for transmission, channel model, propagation loss, ...etc. As a comparative study, we first use Wi-Fi traditional frequency band within the IEEE 802.11 standard and then use the 28GHz mmWave band for the comparison purposes. The network is assumed to have (n) UE at any time and there is a specific number of transmitters and receivers that are willing to exchange data packets at specific times during the network operation. Traditional Wi-Fi UE is assumed to use unidirectional antennas with an equal gain in all directions, whereas the mmWave UE is equipped with directional antennas that can be directed in specific directions with larger gain within these directions. This directionality and antenna gain is the reason behind the different channel models and performance differences reported. More details about the system model will be mentioned in section 5.

### 3.2 Ad Hoc Traditional Routing Algorithms

Many routing and data forwarding algorithms have been proposed for mobile ad hoc networks [8]–[11]. Three of the most famous ones in the literature [9]–[11] are also the ones that we will compare their performance when using sub-6GHz frequencies vs. when using mmWave frequencies. Ad-Hoc On-demand Distance Vector (AODV) is a reactive routing protocol that floods the network with Route Request (Rreq) packets when required [10]. AODV does not rely on periodic advertisements, which reduces the overhead and provide more bandwidth for users. Also, it is proven to be a loop-free routing protocol even in case of mobility and repairing broken links. It scales well with large numbers of mobile nodes that are cooperating to form an adhoc networks. It is expected that using mmWave frequencies with directional antennas with such protocol would improve the overall performance as it will reduce interference (due to directional narrow beams communications) and the large gain the directional antennas can provide toward the relay nodes or the final destinations.

Destination-Sequenced Distance-Vector (DSDV), on the other hand is a proactive (table driven) Routing protocol that is a destination based protocol with no need for a global view of the network topology [9]. Considering each mobile host as a specialized router, this protocol periodically advertises its view of the network topology to other hosts in the network. With such a mechanism, it can modify the Routing Information Protocol (RIP) [22] to be suitable for dynamic and self-starting networks (such as the MANET). Again, using mmWaves with directional antennas and large directed gains can improve such protocol performance due to fewer interference effects and better received SNR at any relay or destination nodes within the transmission range of the mmWave devices.

Finally, Optimized Link State Routing protocol (OLSR) is another table-driven routing protocol for mobile ad hoc networks that exchanges periodic messages to maintain the network topology information at each node [11]. OLSR is an optimized protocol over a pure link state protocol because it compares the size of data sent in each message to reduce the number of retransmissions while flooding the entire network with these messages. It uses multipoint relays technique to efficiently flood the network. Again, using mmWaves with such protocol is expected to improve the performance due to the reduction of interference among these entire message flooding processes and improving the received SNR due to the antenna gain and directionality.

## 3.3 Differences from Traditional MANET Routing Algorithms

We have studied some channel models suggested by 3GPP in [12], and their impact on the performance of some well-known MANET routing protocols [9]–[11]. To understand the difference in performance between traditional Wi-Fi MANET and the networks that use mmWave, we need first to clarify the following:

• Wi-Fi devices broadcast wireless signals in all directions and cover larger distances (up to several miles), whereas mmWave devices only transmit narrow beams in specific directions and cover shorter distances (up to few hundred meters for the Ultra-Dense Networks (UDN) [5].
• Path Propagation loss for Wi-Fi signals is determined by Friis equation as follows (assuming no Transmission gain ($G_t$) or reception gain ($G_r$) or $G_t$ and $G_r = 0$dB):

$$L = 4\pi df/c^2 \tag{1}$$

$$PL(dB) = 20\log(f) + 20\log(d) - 147.56dB \tag{2}$$

Where:
L: is the path loss.
d: is the distance between the transmitter and receiver.
f : is the used frequency.
c: is the speed of light ($3*10^8$) [13].

If we use the same path loss for mm Wave UE and take the Tx gain and Rx gain in consideration (which ranges between 14-17 dB as in [15], then the equation will be:

$$PL(dB) = 20\log(f) + 20\log(d) - 147.56dB - 17dB - 17dB. \tag{3}$$

Which clearly reduces the path loss to a large extent.
Many other path loss models for mm waves have been proposed in the literature recently. According to 3GPP in [14], assuming rural Line of Sight (LoS) path between any two nodes in the MANET, the path loss can be defined as:

$$PL(dB) = 20\log(40\pi df/3) + \min(0.03h^{1.72},10)\log(d) - \min(0.044h^{1.72},14.77) + 0.002\log(h)d. \tag{4}$$

Where:
PL(dB): is the path loss in dB.
d: is the distance between transmitter node and receiver node.
f : is the used carrier frequency.
h: is the height if the Tx node.

The other path loss model of mmWave signals was suggested in [23] and has been proved to match the high fitness of the Line of Sight (LoS) and the Non-Line of Sight (NLoS) environments is the one- parameter close-in (CI) model described below:

$$PL^{CI}(f, d)[dB] = FSPL(f, 1m)[dB] + 10n \log_{10}(d) + X_\sigma^{CI} \tag{5}$$

where n denotes the single model parameter, the path loss exponent (PLE), with 10n describing the path loss in dB in terms of decades of distances beginning at 1m, d is the separation distance between the transmitter and receiver nodes, $X_\sigma^{CI}$ is the SF standard deviation describing large-scale signal fluctuations about the mean path loss over distance, and FSPL(f, 1m) denotes the free space path loss in dB at the transmitter-receiver separation distance of 1m at the carrier frequency f . Also, the free space path loss (FSPL) can be described as:

$$FSPL(f, 1m)[dB] = 20 \log_{10}(4\pi f/c). \tag{6}$$

where c is the speed of light.

It is clearly proven now that there is no one model that is capable of describing the mmWave channel in different environments and that the transmission scenario conditions need to be taken into consideration when trying to talk about such channel models [12], [14]. The following figure shows some of the propagation path loss for different frequencies and distances according to some of the previously described models:

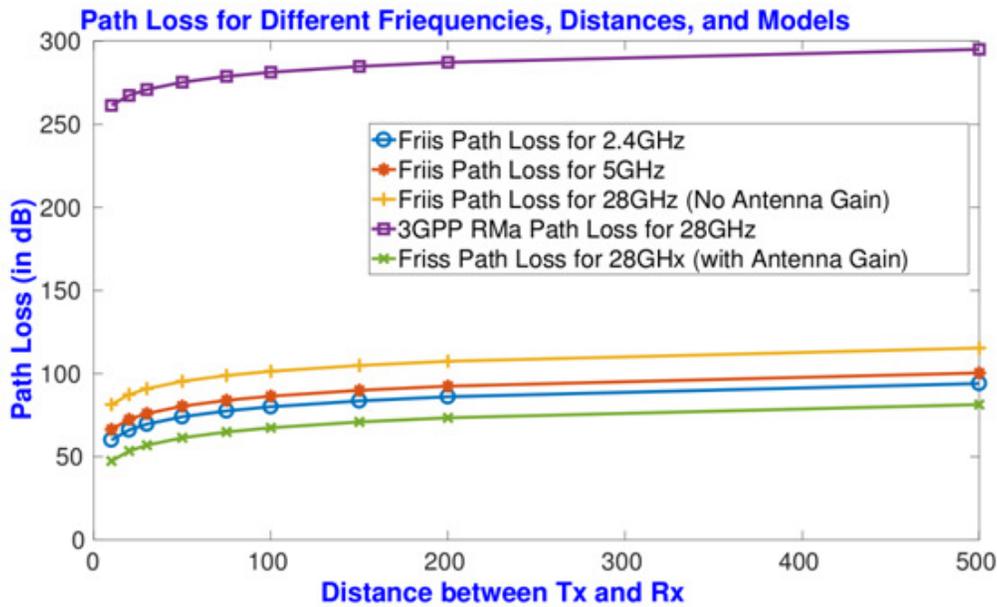

Fig. 1: Path Loss Comparison for different Frequencies, Models, and Distances

The figure above suggests that mmWave devices suffer from more severe path loss than the traditional Wi-Fi devices if we don't take the directional antenna gain in consideration. While this is only true for omnidirectional antennas, using beamforming and directional antennas has been proven to have much less path loss than that of equivalent Wi-Fi devices for short distances [14]. These facts suggests that the UE in MANET that uses Wi-Fi frequencies are tended to transmit messages to farther nodes which means more propagation loss (and more probability of errors in messages and reduction in delivery ratio), whereas the UE with mmwave are tend to transmit messages to closer nodes and with narrow beams with high directional gains which means less propagations loss (and better delivery rates). the reduced propagation loss of UE's with mmwave means better received Signal to Noise (SNR) which leads directly to better delivery ratio as we will see in the next section.

Finally, to examine the superiority of some traditional routing protocols in MANET when using mmwave over their performance when using the sub-6GHz band, we will use a network with specific features and network scenarios and compare these scenarios with:

• Different channel models for Wi-Fi and mmwave signals.
• Different Data rates.
• Different packet sizes.
• Different Tx power.
• Different Routing Protocols.

## 4. PERFORMANCE EVALUATION METRICS

The metrics used to compare the MANET routing algorithms with sub-6GHz band and the ones with mm Wave are listed below:

• Number of Delivered Packets: total number of delivered packets to all destinations per each simulation second.

• Packet Delivery Ratio (PDR): Which is defined as the number of packets received divided by the number of packets sent each second.

$$PDR = Packets_{received}/Packets_{sent} \qquad (7)$$

• Average Delivery Ratio: Which is the mean of all the Packet Delivery Rates for the entire network operation lifetime.

## 5. PERFORMANCE EVALUATION

### 5.1 Network Settings

The goal of this work is to analyze the feasibility of routing protocols in mmWave MANETs, and then compare its performance with the traditional routing protocols in MANET for different network settings. In this section, we show a comparison of three of the most famous routing algorithms in the MANET networks and these are: Destination-Sequenced Distance Vector (DSDV) [9], Ad hoc On-demand Distance Vector (AODV) [10], and Optimized Link State Routing (OLSR) [11] under a typical random waypoint mobility model [6]. The newly proposed module for mmWave in (ns-3) [6] is used for simulation as it provides different channel models for mmWave that are derived from many measurement campaigns done in different places and with different environmental conditions recently. This module that was explained in [6] and [26] with more details focuses on the modeling of the customizable channel, physical and medium access control (MAC) layers of millimeter wave systems and was utilized throughout the simulation steps explained in the next sections.

The comparison is performed first among these protocols in the traditional Wi-Fi frequency range; then we evaluate the same algorithms' performance under different mmWave channel models for the used devices such as the Urban Macro-cells (UMa) and Rural Macro-cell (RMa) as suggested in [12]. The basic simulation scenario in the ns3 runs for 200 simulated seconds, where the first 50 seconds are used for start-up time. The number of UE nodes is 50, and the nodes are moving according to Random Way point Mobility Model with a speed of 20 m/s and no pause time within a 300x1500 m region. The WiFi (which is the basic type of communications in this scenario that we intend to change to higher frequencies) is in ad-hoc mode with a 2 Mb/s rate (802.11b) and a Friis loss model (for Wi-Fi) and 3GPP propagation path loss (for mmWave). The transmit power is set to 7.5, 10, 20, 40 dBm. In this scenario, there are 10 source/sink data pairs sending UDP data at an application rate of 2.048 and 4.96 Kb/s each. This is done at a rate of 4 64-byte and 128-byte packets per second. Application data is started at a random time between 50 and 51 seconds and continues to the end of the simulation.

## 5.2 Simulation Results

Details of simulation scenario parameters are listed in table 1.:

TABLE I: Simulation Scenario Parameters

| Parameters | Specifications |
|---|---|
| OS | Linux Ubuntu 16.04 LTS |
| Network Simulator | ns-3.27 |
| Simulation Time | 200 Seconds |
| Simulation Area | 1500 m X 1500 m |
| Number of Wireless Nodes | 50 |
| Speed of Mobile Nodes | 20 m/s |
| Mobility Model | Random Way Point |
| Data Rate | 2Mbps and 4Mbps |
| Tx Power | 7.5, 10, 20, 40 dBm |
| Number of Tx Nodes | 10 |
| Number of Rx Nodes | 10 |

Some performance comparisons between the traditional Wi-Fi MANET and mm Wave MANET are shown in Figures 2, 3, and 4.

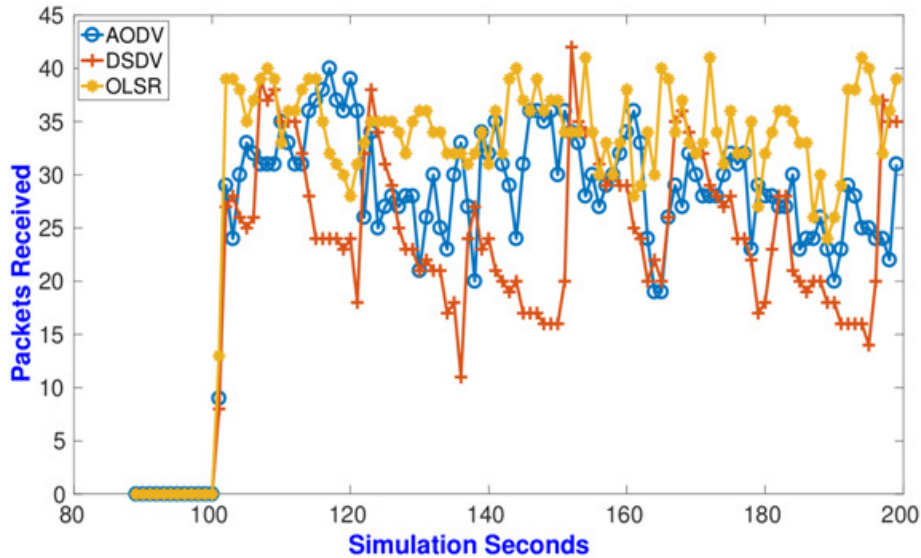

Fig. 2: Packets Received for Different Wi-Fi MANET Routing Protocols

Figure (2) shows the number of received packets each second using AODV, DSDV, and OLSR and it is clear that OLSR is better than the other protocols most of the time for the scenario explained in section 5.1 above.

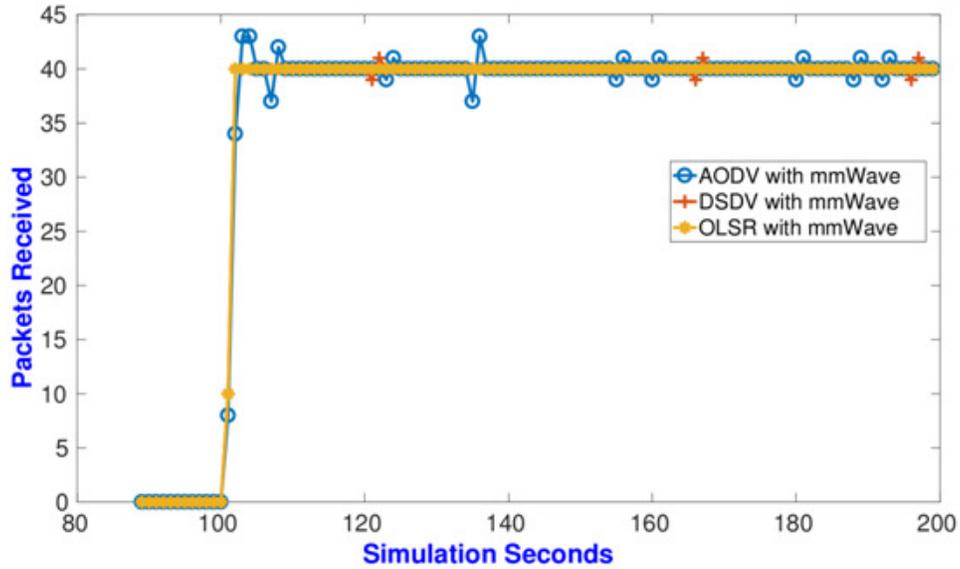

Fig. 3: Packets Received for different mmWave MANET (with RMa Channel Model) Routing Protocols

Figure (3), on the other hand, shows the number of received packets by the same routing protocols when using mmWave frequencies and it is clear that these protocols show more stability and better delivery ratio when used with mmWave than with the traditional sub-6GHz frequencies.

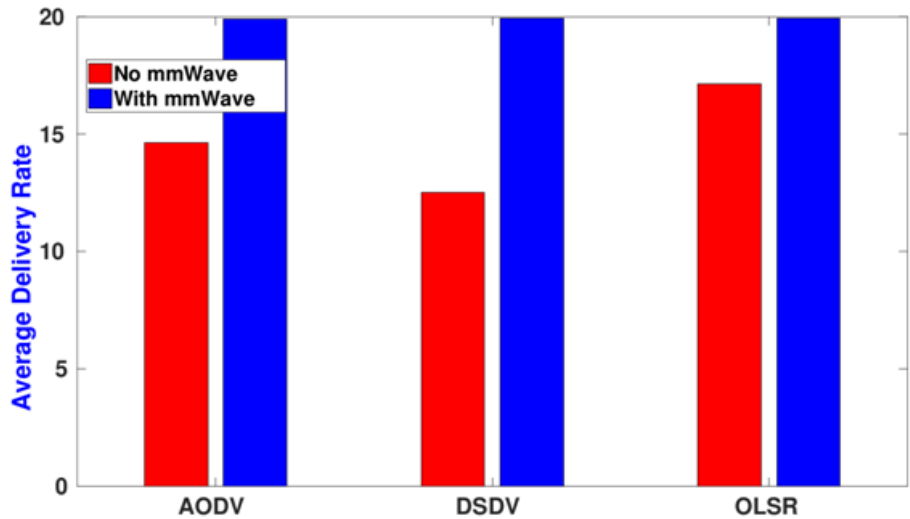

Fig. 4: Average Delivery Ratio for both Wi-Fi MANET and mmWave MANET (with RMa Channel Model) Routing Protocols

Figure (4) is showing a comparison of the average packet delivery ratio between different routing protocols with mmWave versus the same protocols with traditional sub-6GHz frequency bands. The same performance comparisons that were done in figures 2 3 and 4 are repeated for the mmWave

MANET routing protocols under the UMa channel model in the Figures 5 and 6.

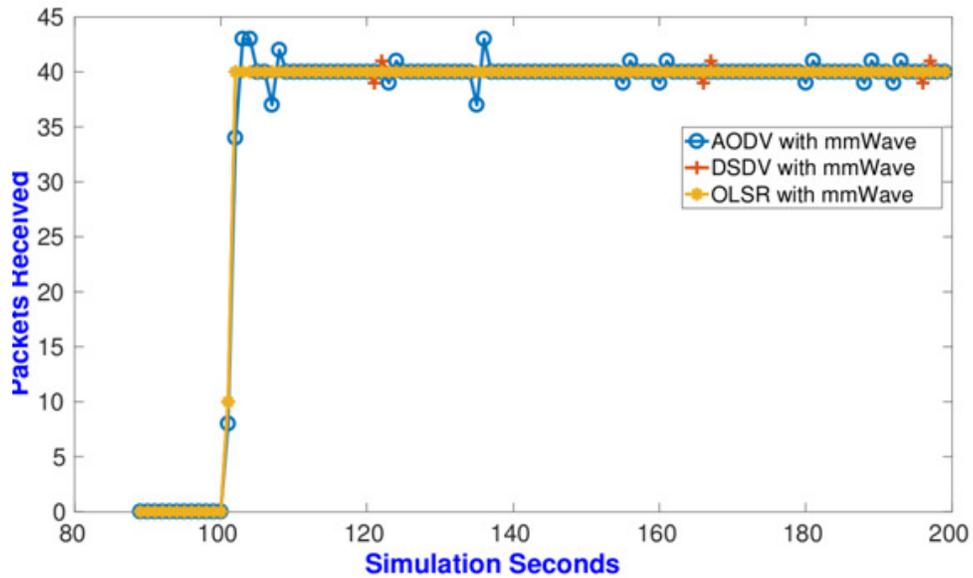

Fig. 5: Packets Received for different mmWave MANET (with UMA Channel Model) Routing Protocols

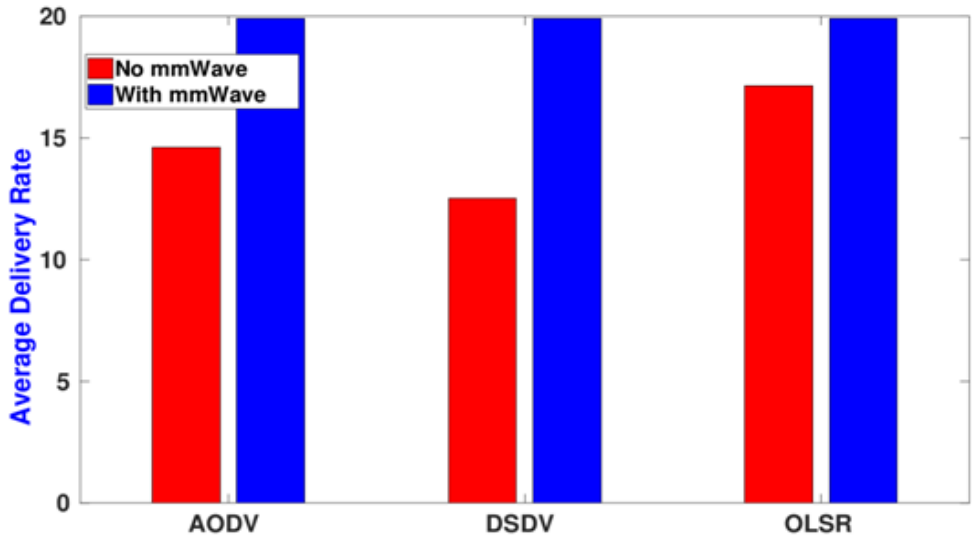

Fig. 6: Average reception rate for different routing protocols for mmWave UMa channel

Now, we change the data rate and packet size of the messages and repeat the comparison in Figures 7, 8, and 9 for the received packets the average delivery rate:

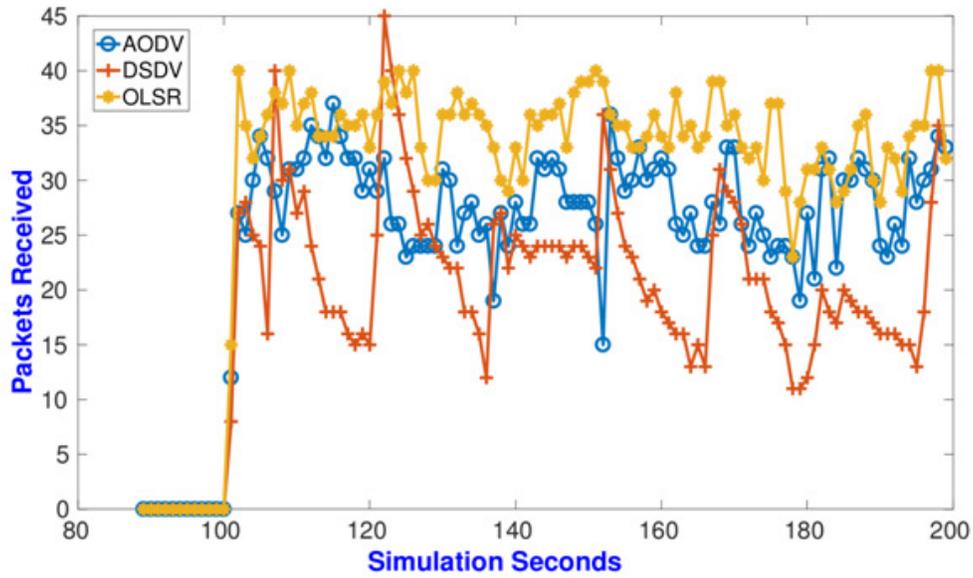

Fig. 7: Packets Received for Different Wi-Fi MANET Routing Protocols with 4kbps transfer rate and 128 Byte packet size

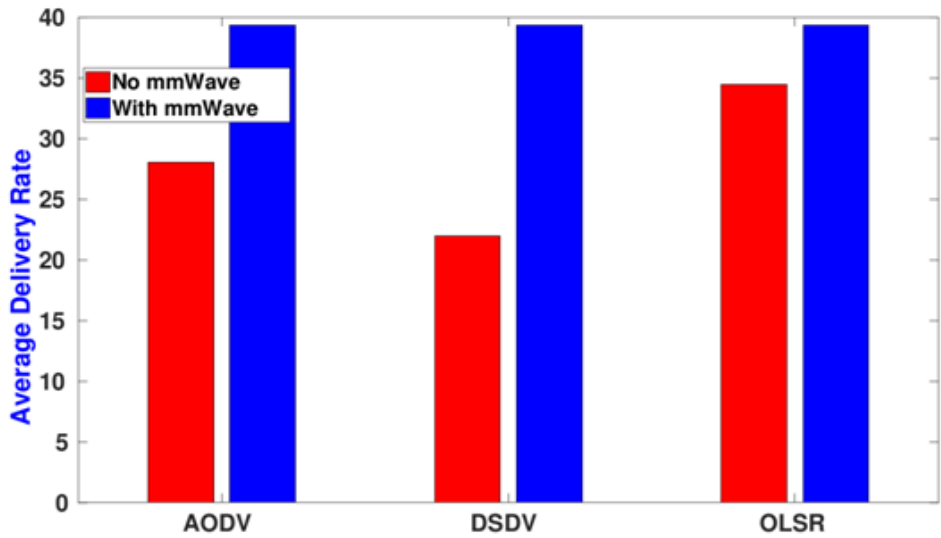

Fig. 8: Average Delivery Ratio for both Wi-Fi MANET and mmWave MANET (with UMa Channel Model) Routing Protocols with 4kbps transfer rate and 128 Byte packet size

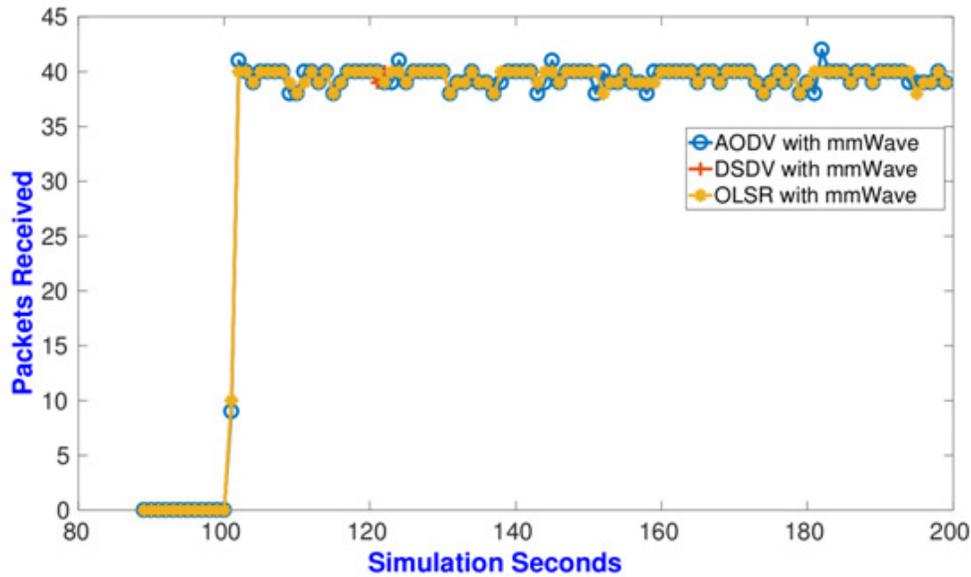

Fig. 8: Packets Received for different mmWave MANET (with UMa Channel Model) Routing Protocols with 4kbps transfer rate and 128 Byte packet size

As can be seen, the delivery ratio and the number of delivered packets during the work of the network in the simulator are much better and more stable for the mmWaves rural and urban channels than the normal Wi-Fi channels. This shows the huge potential for the mmWave in the short-range communications as it is planned in the Ultra Dense Networks (UDN) [5]. More investigations need to be done in this field to unveil the properties and limitations of the mmWave in the MANET field. Also, besides the delivery rate and propagations loss, the effect of large bandwidth (that mmWave brings as a feature) on the network performance and battery-driven devices lifetime (especially in the disastrous regions) much be studied as well.

The final step in our investigation for the mmWave in MANET, is the effect of transmission power on the delivery ratio. It is well known that for broadcast wireless channels, increasing the power would reduce the effect of path loss because of interference and attenuation, but for the directional beams of mmWave, less power should be enough to perform the same. According to the recent FCC in [15], the highest UE effective isotropic radiated power (EIRP) is 43dBm (almost 20 watts). So, we studied the effect of increasing the Tx power of the UE's on the delivery rate of the data packets for different routing protocols in ad-hoc networks and the results shown in figure 10 are as expected, showing that mmWaves are doing better than traditional Wi-Fi frequencies even with less Tx power:

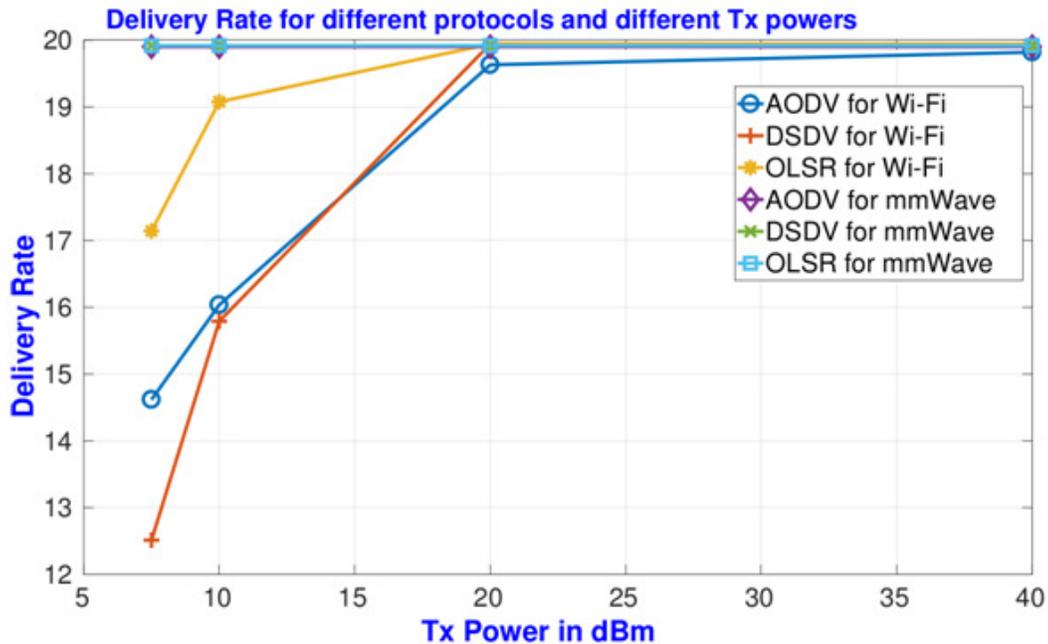

Fig. 10: Tx Power of each UE and its Effect on the Delivery Ratio

For the Tx power effect on the MANET routing protocols performance, we can see that increasing the power helped in increasing the delivery ratio of the Wi-Fi MANET routing protocols, whereas it did not help in the case of mmWave as the delivery ratio was almost constant, but it is still better than that of the Wi-Fi networks. This can be explained by taking in consideration that mmWave devices with the help of beamforming utilize the power better than the Wi-Fi devices and concentrate the Tx power in a narrow beam to reduce the path loss. This means that the mmWave devices can prolong the network lifetime as they provide energy efficiency and reduce the power consumption compared with the Wi-Fi devices to cover the same area and provide even better performance.

## 6. Conclusions And Future Work

In this paper, we studied the feasibility of some well-known routing protocols for mobile ad-hoc networks with mmWave frequency bands and showed how utilizing mmWave frequencies can increase the network efficiency and delivery ratio. Several parameters of the network have been adjusted and in each case the MANET with mmWave was shown to be better than the Wi-Fi counterpart. Simulation using mmWave module of the ns-3 simulator (that was developed and released recently [6], [26]) was used to confirm the results. Further investigation of the utilization of mmWave frequencies in different types of Self-Organized Networks (SON) by utilizing the unique features the mmWave frequencies offer for such networks is part of our future work. Also, new routing protocols for such networks that depend solely on the random beamforming data forwarding is yet to be studied.

## AUTHOR

Mustafa S. Aljumaily received his MSc in Computer Engineering from the University of Basrah, in 2010. He is currently a PhD candidate in the Electrical Engineering and Computer Science Department, University of Tennessee, Knoxville, USA. His research interests include Human Machine Interaction (HMI), Gestures recognition, Wireless Sensor Networks (WSN), Millimeter Wave Wireless communications (mmWave), 5G, Self-Organized Networks (SON), Opportunistic Routing in Mobile Ad-Hoc Networks (MANET). He has published papers in several conferences and journals including  IJPCC and IJCSI.

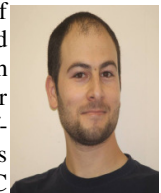